\documentclass[twocolumn,showpacs,preprintnumbers,amsmath,amssymb,floatfix]{revtex4}

\usepackage{graphicx}
\usepackage{dcolumn}
\usepackage{bm}

\begin{document}
\newcommand*{\bra}[1]{\ensuremath{
  \left\langle #1 \right\vert }}
\newcommand*{\ket}[1]{\ensuremath{
  \left\vert #1 \right\rangle }}
\newcommand*{\braket}[2]{\ensuremath{
  \left\langle #1 | #2 \right\rangle}}
\newcommand*{\E}[1]{\ensuremath{
  \left\langle #1 \right\rangle}}
\newcommand*{\partiald}[2][]{\ensuremath{
  \frac{\partial #1}{\partial #2} }}
\newcommand*{\totald}[2][]{\ensuremath{
  \frac{d#1}{d#2} }}
\newcommand*{\vect}[1]{\ensuremath{
  \mathbf{#1}}}
\newcommand{\displacement}[2][]{\ensuremath{
  \hat D^{#1} \negmedspace \left( \left\{ #2 \right\}\!\right)}}

\preprint{APS/123-QED}

\title{Continuous measurement with traveling wave probes}

\author{Andrew Silberfarb}
\author{Ivan H. Deutsch}
 \email{ideutsch@info.phys.unm.edu}
\affiliation{Department of Physics and Astronomy, University of
New Mexico, Albuquerque, NM 87131}

\date{\today}

\begin{abstract}
We consider the use of a traveling wave probe to continuously
measure the quantum state of an atom in free space.  Unlike the
more familiar cavity QED geometry, the traveling wave is
intrinsically a multimode problem.  Using an appropriate modal
decomposition we determine the effective measurement strength for
different atom-field interactions and different initial states of
the field.  These include the interaction of a coherent-state
pulse with an atom, the interaction of a Fock-state pulse with an
atom, and the use of Faraday rotation of a polarized laser probe
to perform a QND measurement on an atomic spin.
\end{abstract}


\maketitle

\section{\label{S:Introduction} Introduction}

Current research to create quantum information processors has
forced a re\"{e}xamination of the underlying description of these
devices.  In order for us to gain information about quantum
systems they must be measured.  Whereas the standard picture of
such measurements involves the ``collapse of the wave function"
following Von Neumann's projection postulate, such strongly
measuring probes are rarely implemented in the laboratory.  More
typically, a continuous probe interacts with the system which is
then detected as a macroscopic signal. Examples include the
probing of a quantum dot with a single electron transistor (SET)
\cite{Devoret2000} and the measurement of the position of a
micro-mechanical cantilever by monitoring the modulation of a
reflected laser beam \cite{Milburn1994}.  The formalism of quantum
mechanics provides a number of different approaches for analyzing
such situations.  Scattering theory employs Green's function
input-output relations to describe the evolution of the probe
asymptotically, both before and after its interaction with the
system under examination.  Alternatively, the theory of quantum
trajectories \cite{Carmichael1993} provides a dynamical
description of the quantum system being measured, conditioned on
the continuous information being collected via the probe.  As
laboratory developments give us access to the control and
manipulation of quantum systems these descriptions become ever
more relevant. The quantum trajectory approach has the advantage
of directly tying the dynamics of the system's evolution to the
measurement record.  The ability to do this is essential when
implementing adaptive measurement and control strategies employing
feedback \cite{Wiseman1994, Doherty1999,Smith2002,Morrow2002}.

The standard paradigm for continuous measurement is cavity QED.
The dynamics of a cavity mode of the electromagnetic field
(perhaps coupled to an atom) are monitored by a partially
transmitting mirror \cite{Mabuchi1999,Foster2000}.  Input-output
scattering theory, suited specifically to the language of optical
elements \cite{Gardiner1985}, is used to connect the intracavity
dynamics with those of the traveling signal.  In order to
translate the typically discrete information of the individually
transmitted photons into continuous information one considers a
homodyne or heterodyne measurement, in which the signal is mixed
with a macroscopic local oscillator.  The result is a stochastic
Schr\"odinger equation which describes both the localization of
the quantum state conditioned on the measurement and also the
effect of quantum ``back-action noise" \cite{Gardiner2000}.

In many applications, there is no confining cavity to set the
interaction strength between the system under observation and the
probe.  Though a ``quantization volume" may be employed as a
calculational tool, its imagined characteristics should not
determine physically relevant quantities.  For example, Milburn
{\em et al.}\ \cite{Milburn1994} modelled continuous observation
of a moving cantilever monitored by the modulation of a reflected
laser beam. They considered the cantilever to be a mirror, which,
partnered with an imagined partially transmitting surface, formed
a leaky optical cavity.  Under the assumption that the
transmission rate of light through the fictitious mirror is much
faster than the characteristic rate at which the cantilever moves,
the cavity could be adiabatically eliminated from the dynamics.
This led to a stochastic Schr\"odinger equation for the
continuously observed cantilever alone. The unphysical
quantization volume, however, still appeared implicitly in this
equation.

The key parameter that characterizes the dynamics of a
continuously observed system is the ``measurement strength", which
we will denote $\kappa$.  It determines the rate at which
information is gathered about the system and consequently sets the
scale at which effects such as quantum back-action become
significant.  For example, Bhattacharya et al.\
\cite{Bhattacharya2000} have shown that for sufficiently
macroscopic systems there is a window of values for $\kappa$ such
that continuous observation can localize the probability
distribution to a quantum trajectory that faithfully tracks the
classically predicted trajectory, with minimal quantum noise.
Another example is the continuous measurement of ensembles of
atoms, controlled through their collective interaction with a
common probe, to produce nonclassical spin squeezed states
\cite{Kuzmich2000, Julsgaard2001}.  These effects depend crucially
on $\kappa$ and its relation to the other rates governing the
system dynamics.

In this article we formulate the problem of continuous measurement
by a traveling wave probe.  We derive a master equation describing
the situation in which the system is monitored by the probe, but
the measurement result is not recorded.  This allows us to to
identify the important characteristic scales of the problem
without reference to a particular measurement scheme.  We begin in
Sec.\ \ref{S:Propagation}  by establishing the necessary formalism
for treating propagating fields in quantum optics, in contrast to
the more familiar closed cavity problems. We apply this formalism
in Sec.\ \ref{S:Interaction} to the fundamental problem of a
two-level atom interacting with a resonant laser field.  When the
field is treated classically this leads to Rabi flopping, but when
treated quantum mechanically the laser not only manipulates the
atom but also acts to continuously measure it.  We determine the
rate at which the measurement back-action leads to decoherence in
the atomic system. This is contrasted with the evolution when the
atom is coupled to a resonant electromagentic pulse with a fixed
photon number $n$. In particular, we explore the circumstances
under which we can recover the usual Jaynes-Cummings solution for
a two-level atom coupled to a single mode \cite{Scully1997} and
show how the behavior for a single photon diverges from this
solution. In Sec.\ \ref{S:Faraday} we consider continuous
measurement of an atomic spin through the Faraday rotation of an
off-resonant laser field. This process has been used to generate
spin-squeezed states in atomic ensembles \cite{Kuzmich2000,
Julsgaard2001}. We conclude and summarize our results in Sec.\
\ref{S:Summary}.

\section{\label{S:Propagation} Quantum description of propagating fields}

Classically, when considering quasimonochromatic propagating
fields, it is natural to model the evolution of the system as a
function of the propagation direction, $z$. The field at $z$ can
then be decomposed into a complete set of orthonormal temporal
modes which act locally. One might be tempted to describe the
quantum fields in an analogous manner by quantizing the temporal
modes at fixed position, $[\hat a(t),\hat a^\dag (t')]=\delta
(t-t')$. The field operator could then be decomposed into a
complete set of orthonormal mode functions, $\phi_i(t)$, so that
$\hat a(t)=\sum_i {\phi _i(t)\,\hat c_i}$, with $[\hat c_i, \hat
c^\dag_j]= \delta_{ij}$. Boundary conditions at some initial plane
could then used to restrict the mode content, possibly to a single
temporal mode.

This approach was taken by van Enk and Kimble \cite{vanEnk2001a}
and also by Gea-Banacloche \cite{Banacloche2002} who considered an
analogous problem to the one we address here . They studied how
errors were generated in quantum logic operations due to the fact
that control pulses are not truly classical and can become
entangled with the atoms with which they are interacting. Their
analysis led to an effective single temporal mode theory. Though
some of their conclusions are correct, one must take great care to
understand the regimes under which this formalism is applicable.
Consider, for example, a single photon pulse interacting with a
localized two-level atom.  Let us suppose that the duration of the
pulse is short compared to the natural lifetime of the atom in its
excited state but sufficiently long to be considered
quasimonochromatic and on resonance.  Defining creation and
annihilation operators for the single temporal mode associated
with this pulse, the Hamiltonian, in the rotating wave
approximation, appears to have the familiar Jaynes-Cummings form,
\begin{equation}
    \hat H=\hbar g\left( \hat a \hat \sigma _+ + \hat a^\dag \hat\sigma _- \right),
\end{equation}
where $\hat\sigma_{\pm}$ are the usual raising and lowering
operators associated with the two-state atom. Given the atom
initially in its ground state, the solution leads to quantum Rabi
oscillations,
\begin{equation}\label{E:Oscillation}
  \ket{\psi (t)}=\cos(gt)\,\ket{g}\ket{1}
    -i\sin(gt)\,\ket{e}\ket{0}.
\end{equation}
This falsely predicts the possibility of a single photon
$2\pi$-pulse in free space, whereby the photon is perfectly
absorbed and then re\"{e}mitted into the original mode.  In
reality once the atom has absorbed the photon it will re\"{e}mit
into a mode consistent with its radiation pattern, \textit{not}
into the initial packet mode. That is, the single photon will be
\textit{scattered}. This emission must also obey causality; no
information about the emitted photon can register on a distant
detector at a space-like separated point. In the solution above,
however, the atom both absorbs from and emits into a spatially
delocalized photon mode in free space, violating causality.

The problems with causality arise from the faulty quantization
procedure outlined above.  Quantum fields must be defined over all
space at an initial time (more generally on an initial space-like
surface).  Unitarity then ensures that \textit{equal-time}, not
\textit{equal-space}, commutation relations are preserved.
Nonequal-time commutation relations cannot generally satisfy the
canonical commutation relations, being inconsistent with
Poincar\'e invariance \cite{Bjorken}.  The exception is for free
fields, or fields that behave like them (e.g.\ fields traveling
through matter whose response is approximately linear, or
asymptotic ``in" and ``out" fields as used in scattering theory).

We review here a formalism appropriate for treating the quantum
optics of paraxial propagating fields \cite{Deutsch1991a}.
Consider a quasimonochromatic paraxial beam with frequency
$\omega_0$ and wave number $k_0$. We write the positive frequency
component,
\begin{equation}
  E^{(+)}(\vect{x},t)=\,\mathcal{E}(z,t)\,\phi_T(x,y)e^{i(k_0z-\omega_0t)},
\end{equation}
where $\exp [i(k_0z-\omega _0t)]$ is the ``carrier wave", $ \phi
_T(x,y)$ is the ``transverse mode" (e.g.\ Gaussian), and
$\mathcal{E}(z,t)$ is the slowly varying envelope, meaning its
spatio-temporal variation is much slower than the carrier
wavelength/frequency.  We have ignored both diffraction and the
vector nature of the field.  It is easy to show that the free
space wave equation becomes,
\begin{equation}
  \left( \partiald{t}+c \partiald{z} \right)\mathcal{E}(z,t)=0
\end{equation}
for the envelope, whose solution is
$\mathcal{E}(z,t)=\mathcal{E}(z-ct,0)$, i.e.\ propagation of the
pulse envelope.  We quantize by replacing the field envelope with
a scaled operator,
\begin{equation}
  \mathcal{E}(z,t)\Rightarrow \sqrt {2\pi \hbar \omega _0}\,\hat\Psi(z,t),
\end{equation}
which satisfies the canonical equal-time commutation relation
\cite{Deutsch1991a},
\begin{equation}
  \left[ {\hat \Psi (z,t),\hat \Psi ^\dag (z',t)}\right]=\delta(z-z').
\end{equation}
This commutator is equivalent to that of a nonrelativistic massive
Bose gas in one dimension.  Here the carrier wave plays the role
of the rest mass and the slowly varying envelope plays the role of
small fluctuations around the mass shell.  The free field
Hamiltonian (removing the carrier wave energy) takes the form,
\begin{equation}
  \hat H_F=c\int dz\,\hat \Psi ^\dag (z)\left( -i\hbar \partiald{z} \right)\hat \Psi (z),
\end{equation}
whose Heisenberg equation of motion gives the wave equation above.
This Hamiltonian is nothing but the second quantized version of
the energy of a photon $E=cp$.

\begin{figure}[t]
  \scalebox{.45}{\includegraphics{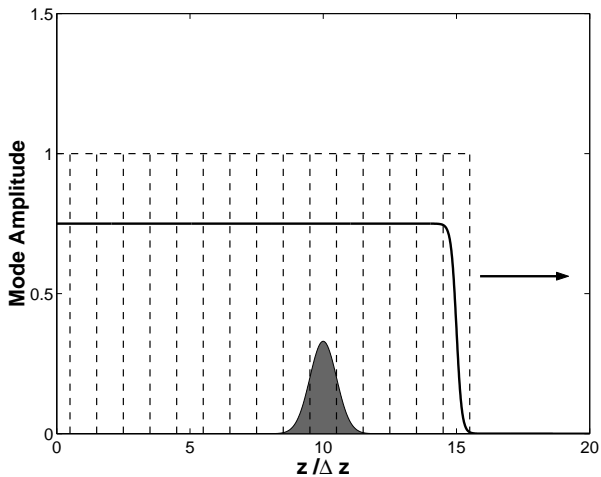}}
  \caption{\label{F:Minimodes} An initial traveling wave pulse (solid line) is
  broken up into many smaller modes $\{\phi_i(z)\}$ (dashed line).  The wave packet of the
  interacting atom (gray) has the
  same width as the mode functions ($\Delta z$).}
\end{figure}

Consider an atom interacting with the field.  In the
electric-dipole and rotating wave approximation the interaction
Hamiltonian is,
\begin{equation}
  \hat H_{AF}=\int d^3x\,\,\left| \Phi (\vect{x}) \right|^2d
    \left( \hat{E}^{(+)}(\vect{x})\hat \sigma _++\hat E^{(-)}(\vect{x})\hat \sigma _-\right),
\end{equation}
where $d$ is the dipole matrix element and $\Phi(\vect{x})$ is the
atom's center of mass wave function.  We take the atom to be
trapped, having center of mass wave function  $\Phi
(\vect{x})=f_T(x,y)\,f_L(z)$ \cite{Coherence}. Then the
interaction Hamiltonian becomes,
\begin{gather}
  \hat H_{AF}=d\sqrt {\frac{2\pi \hbar \omega _0}{A}}
    \int dz\,\left| f_L(z) \right|^2
    \left( \hat \Psi (z)\hat \sigma _+
      + \hat \Psi ^\dag (z)\hat \sigma _- \right) \nonumber\\
  \text{where } \int dxdy\left| f_T(x,y) \right|^2\phi_T(x,y)
    \equiv \frac{1}{\sqrt{A}} ,
\end{gather}
$A$ being the effective area of the mode interacting with the
atom.  Let us go to the interaction picture by including the free
evolution of the atom and field in the interaction Hamiltonian.
Assuming the carrier wave is on resonance,
\begin{align}
  \hat H_{AF}(t)=d \sqrt{\frac{2\pi \hbar \omega _0}{A}}
    \int dz\,\left| f_L(z) \right|^2
    &\left(  \hat \Psi (z-ct)\hat \sigma _+ \right. \\
    &\ \left. + \hat \Psi ^\dag (z-ct)\hat \sigma _- \right).
  \nonumber
\end{align}
Finally, given a set of orthonormal functions (`` longitudinal
modes") $\left\{ \phi _i(z) \right\}$, chosen to be real without
loss of generality,
\begin{equation}\label{E:Split}
  \hat \Psi (z)=\sum_i \phi_i(z)\hat a_i,
\end{equation}
\begin{multline}\label{E:ModeHam}
  \hat H_{AF}(t)=d \sqrt{\frac{2\pi \hbar \omega _0}{A}} \sum_i
    \int dz\,\left| f_L(z) \right|^2 \\
      \phi_i(z-ct)\left( \hat a_i \hat\sigma _+
        + \hat a_i^\dag \hat \sigma_-\right) .
\end{multline}
The Hamiltonian in Eq.\ (\ref{E:ModeHam}) describes the
interaction of each longitudinal mode as it propagates past the
atom(Fig.\ \ref{F:Minimodes}).  Assuming this time scale is much
shorter than any other dynamical scale in the problem, it is
appropriate to make the Markov approximation, whereby one coarse
grains over the time scale over  which the system and reservoir
retain ``memory". To this end, we break up the $z$-axis into
slices of size $\Delta z$, each the extent of the atom wave packet
(e.g.\ the rms of the probability density). The Markov
approximation will hold if the transit time of the field across
the atomic packet, $\Delta t=\Delta z/c,$ is much smaller than any
time scale over which the atom changes. This is certainly an
excellent approximation.  Since we will not consider dynamics on a
time scale smaller than $\Delta t$, we can approximate the set of
atomic wavepackets, centered at each coarse grained slice, as a
complete orthonormal set.  That is, each slice is an approximate
delta function.  Then normalization of both the atomic wave packet and
mode functions combine to give
\begin{subequations}
\begin{eqnarray}
	\int dz\,\left| f_L(z) \right|^2 \phi_i(z-ct) = \frac{1}{\sqrt{\Delta z}} \Theta_i(t)\\
	 \Theta _i(t)=\begin{cases}
    1,&(i-1)\Delta t<t\le i\Delta t \\
    0,& \text{otherwise}
  \end{cases} \label{E:Step}
\end{eqnarray}
\end{subequations}

Under this approximation the Hamiltonian takes the form,
\begin{eqnarray}
  \hat H_{AF}(t)=\hbar g\sum_i \Theta _i(t)\left( \hat a_i\hat \sigma _+
    + \hat a_i^\dag \hat \sigma _- \right)\label{E:Hamiltonian}\\
   \text{where } \hbar g=dE_{\text{vac}}
    =d\sqrt {\frac{2\pi \hbar \omega _0}{Ac\Delta t}} \nonumber
\end{eqnarray}
This result has a clear interpretation.  The traveling wave
configuration is effectively multimode, with each member of the
set being a traveling packet ``mode-matched" to the atom.  The
coupling constant $g$ depends on this mode volume. This picture is
equivalent to a model of decoherence discussed by Brun
\cite{Brun2002} in which a ``flying qubit" passes over a ``system
qubit", the former acting as an irreversible reservoir (through
its continuous spatial degrees of freedom) to carry information
away from the system, thereby leading to decoherence.  In our
problem a given harmonic oscillator (mode of the electromagnetic
field) flies over the qubit, becomes entangled with it, and then
flies away.  This too leads to decoherence, as we describe in the
next section.

\section{\label{S:Interaction} The Rabi interaction for traveling waves}

\subsection{\label{s:Laser} Interaction with a laser beam}

We consider first the case of a resonant laser beam interacting
with a trapped two-level atom.  The state of the field is
described by a tensor product of identical coherent states for
each traveling mode packet,

\begin{equation}
  \ket{\psi}_{\text{beam}}= \bigotimes_i \ket{\alpha}_i ,
\end{equation}
where the amplitude is given by the mean number of photons in time
slice $\Delta t$, $\left| \alpha  \right|^2=P\Delta t/(\hbar
\omega _0)$, with $P$ the power of the laser beam. More precisely,
the state of the beam is a statistical mixture of states of the
form above, averaged over the common, but unknown, phase of the
complex amplitude $\alpha$, as described by van Enk and Fuchs
\cite{vanEnk2002}.  The actual value of the phase plays no role in
the analysis to follow, so we choose it to be fixed with no loss
of generality. In order to distinguish coherent effects from
decoherence we transform by a unitary displacement of the field
states to the vacuum,

\begin{equation}
  \ket{\Psi} \Rightarrow \displacement[-1]{\alpha} \ket{\Psi},
  \quad \hat A\Rightarrow \displacement[-1]{\alpha}
    \hat A \, \displacement{\alpha}.
\end{equation}
In this picture,
\begin{align}\label{E:AVHam}
  H_{AF}(t)&=\hbar g \alpha \left( \hat \sigma _+ + \hat\sigma_-\right)
    + \hbar g\sum_i \Theta _i(t)\left(\hat a_i\hat \sigma _+
      + \hat a_i^\dag \hat \sigma _-\right)\nonumber\\
&=H_{\text{coh}}+H_{\text{AV}} .
\end{align}
The coherent term is classical Rabi flopping at a frequency
$\Omega =2g\alpha =d\sqrt {8\pi I / \hbar^2 c}$, with $I$ the beam
intensity (cgs units, as used throughout).  The second term is the
atom-vacuum coupling for the traveling wave modes only (i.e.\ the
paraxial modes of the beam) \cite{Itano2002}.

We can now proceed with the standard Markov analysis to derive the
Master equation. The initial atom-vacuum state is uncorrelated.
After a time $\Delta t$, one of the modes becomes entangled with
the atom through the atom-vacuum coupling.  The Linblad (``jump")
operator, $\hat{L}$, is defined by \cite{Brun2002},

\begin{equation}
  \bra{1} \hat U_{\text{AV}}(\Delta t)\ket{0}
    =\frac{-i}{\hbar} \Delta t \bra{1} \hat H_{\text{AV}}\ket{0}
    = \hat L \sqrt{\Delta t}.
\end{equation}
Plugging in the Hamiltonian from Eq.\ (\ref{E:AVHam}), we arrive
at,
\begin{eqnarray}
  &\hat L
    =g \sqrt{\Delta t}\,\hat \sigma _-
    =d\sqrt{\frac{2\pi \hbar \omega _0}{Ac}}\,\hat \sigma _-
    \equiv \sqrt \kappa \,\hat \sigma _- \label{E:MeasStr}\\
  & \text{where } \kappa =d^2\left( \frac{2\pi k_0}{\hbar A}\right)
    =\Gamma \left( \frac{3\pi }{2k_0^2A} \right)
         \equiv \Gamma \left( \frac{\sigma_{\text{eff}} }{A} \right)
         ,\nonumber
\end{eqnarray}
is the measurement strength, $\Gamma$ the spontaneous emission
rate and $\sigma_{\text{eff}}$ the effective cross section for
scattering out of the paraxial modes.  By determining these jump operators, we compactly derive the master equation, equivalent to that obtained using a the usual system-reservoir approach after tracing over the unmeasured bath \cite{Scully1997, Gardiner2000, Preskill1998},

\begin{align}
  \totald[\hat{\rho}]{t}&=\frac{-i}{\hbar} \left[ H_{\text{coh}},\hat{\rho} \right]
    -\frac{1}{2} \left\{ \hat L^\dag \hat L, \hat{\rho}  \right\}
    +\hat L\hat{\rho} \hat L^\dag \nonumber \\
  &=\frac{-i}{\hbar} \left[ H_{\text{coh}},\hat{\rho} \right]
    -\frac{\kappa}{2} \left\{ \hat \sigma _+\hat \sigma _-,\hat{\rho}  \right\}
    +\kappa \,\hat \sigma _- \hat{\rho} \hat \sigma _+ .
\end{align}
This equation has a familiar and appealing form.  It is none other
than the master equation for a decaying laser-driven atom
\cite{Scully1997}, but with $\Gamma \to \kappa $.  The decay rate
is due to the entanglement between the atom and the laser modes.
Note, $\kappa$ is independent of $\Delta t$, which acts as a
fictitious ``quantization volume" and so must be absent from any
physical quantities such as the measurement strength.

The measurement strength is also independent of the laser power
$|\alpha|^2$. In particular we may turn off the laser ($\alpha \to
0$), and the measurement strength will remain the same.  The ratio
$\kappa/\Gamma = \sigma_{\text{eff}} /A$ may thus be interpreted
as the fraction of spontaneous emission into the paraxial modes.
In support of this interpretation note that the mode area $A$ can
never be made smaller than the diffraction limit $A\sim 1/k_0^2$,
so at most $\kappa \sim \Gamma $. Moreover, once the beam becomes
focused to such a small spot size, one can no longer neglect the
vector nature of the atom field coupling which further decreases
the measurement strength \cite{vanEnk2001b}.

From Eq.\ (\ref{E:MeasStr}) we can determine how continuous
measurement by the laser beam acts to decohere the atom.  For a
paraxial beam we require that $\sigma_{\text{eff}} /A \gg 1$, so
that diffraction effects are minor. Then decay due to entanglement
with the laser modes is small compared to decay due to spontaneous
emission into $4 \pi$ steradians.  In agreement with the
conclusions of \cite{vanEnk2001a, Banacloche2002}, errors in
coherent control pulses due to the quantum nature of the
interaction can be neglected if spontaneous emission is also
negligible during the duration of the interaction.

\subsection{\label{s:Single} Interaction with a single photon}
In Sec.\ \ref{S:Propagation} we showed how a quantization
procedure in terms of nonequal-time commutators can lead to a
false prediction of single photon coherent Rabi flopping in free
space. In this subsection we use our formalism to show how a
quasimonochromatic and paraxial propagating single photon wave
packet drives a two-level atom.

We take the initial state of the system to be a single photon wave
packet with the atom in its ground state,
\begin{equation}
  \ket{\Psi (0)} = \hat a^\dag[f] \ket{\text{vac}}  \otimes\ket{g}_{\text{A}}.
\end{equation}
The operator $\hat a^\dag[f] = \int dz \hat \Psi^\dag (z) f(z)$
creates a delocalized single photon state with slowly varying
pulse envelope $f(z)$ \cite{Deutsch1991b}.  For simplicity we use
a square pulse of duration $\tau = N \Delta t$.  In this case we can 
expand $f(z)$ in a symmetric sum of
coarse-grained modes each having length $\Delta z = \Delta t/c$,
\begin{equation}\label{E:SingleInit}
   \hat a^\dag[f] = \frac{1}{\sqrt{N}} \sum_{i=0}^{N-1} \hat a_i^\dag.
\end{equation}
The state will evolve according to the Hamiltonian in Eq.\
(\ref{E:Hamiltonian}) which commutes with the total number of
excitations in the system. Neglecting, for now, the possibility of
spontaneous emission into other transverse field modes, the total
number of excitations will be preserved. The state at all times
must then have the form
\begin{equation}\label{E:Expansion}
  \ket{\psi(t)} = \left(\sum_j A_j(t) \hat a_j^\dag + A_e(t) \hat \sigma_+ \right) \ket{\text{vac},g}.
\end{equation}

Consider the evolution of the system over the short interval $(t_k
,t_k + \Delta t]$, where $t_k = k \Delta t, 0 \leq k \leq N-1$. We
can define a map for the state between two successive time steps,
\begin{equation}
  \ket{\psi(t_k)} = e^{-i \hat H_k \Delta t /\hbar}
    \ket{\psi(t_{k-1})},
\end{equation}
where $\hat H_k =  \hbar g \left(\hat a_k \hat \sigma_+  + \hat
a^\dag_k \hat\sigma_- \right) .$  Using ansatz (\ref{E:Expansion})
we are led to the recursion relations,
\begin{subequations}
\begin{align}
  A_{j\neq k}(t_{k}) &= A_{j}(t_{k-1}),\label{E:Recursion1}\\
  A_{k}(t_{k}) &= A_k(t_{k-1})c -i A_e(t_{k-1}) s, \label{E:Recursion2}\\
  A_{e}(t_k) &= A_e(t_{k-1})c -i A_k(t_{k-1})s. \label{E:Recursion3}
\end{align}
\end{subequations}
Here $s \equiv \sin \sqrt{\kappa \Delta t}$ and $c \equiv \cos
\sqrt{\kappa \Delta t}$.  The measurement strength $\kappa = g
\sqrt{\Delta t}$ is the same as in Eq.\ (\ref{E:MeasStr}).

These coupled algebraic equations can be solved for the
amplitudes.  Repeated application of the Eq. (\ref{E:Recursion1})
at all times $t_k < t_j$, shows that $A_k(t_{k-1}) = A_k(0) =
N^{-1/2}$. Inserting this result into Eq. (\ref{E:Recursion3}),
\begin{equation}\label{E:Difference}
  A_e(t_k) = A_e(t_{k-1})c - \frac{i s}{\sqrt{N}}.
\end{equation}
This equation admits a simple series solution,
\begin{equation}
  A_e(t_k) = -\frac{is}{\sqrt{N}} \frac{1-c^k}{1-c}.
\end{equation}
Assuming $\sqrt{N} \gg 1$, i.e.\  the envelope $f(z)$ is much
broader than the coarse graining, we may take the limit of
$A_e(t)$ as $N\to \infty$ holding $\tau =N \Delta t$.  This
yields,
\begin{equation}\label{E:NoSpont}
  A_e(t) \approx -\frac{2 i}{\sqrt{\kappa \tau}}
    \left[1 - e^{- \kappa t/2}\right].
\end{equation}

The solution given in Eq.\ (\ref{E:NoSpont}) is based on the
fundamental assumption that the evolution of the state is unitary,
i.e.\ we consider a closed  quantum system consisting of the atom
and paraxial field modes.  In the continuum limit, this yields an
effective decay due to emission into the included paraxial modes
at rate $\kappa$, but it excludes decay into all others modes
which, taken together, give a total spontaneous emission rate
$\Gamma$.  Since we showed in Sec.\ (\ref{s:Laser}) that $\kappa
\ll \Gamma$, this solution is not self-consistent.  To rectify
this, during the time interval $\Delta t$ we must allow for a
small probability of  spontaneous emission, $P_\text{spont}$, into
non-paraxial modes. By not including these modes in our system,
the state ket $\ket{\psi}$ evolves according to an effective
non-Hermitian Hamiltonian \cite{Gardiner2000} with decaying norm,
$\braket{\Psi}{\Psi} = 1- P_\text{spon}$.  Eq.\
(\ref{E:Recursion3}) then reads,
\begin{equation}
  A_{e}(t_k) = A_e(t_{k-1})c e^{-\gamma \Delta t/2} - i A_k(t_{k-1}) s,
\end{equation}
where $\gamma$ is the spontaneous emission into all non-paraxial
modes.  Employing the initial condition and taking the limit
$\Delta t \to 0, N \Delta t = \tau$ yields,
\begin{equation}\label{E:SpontDiff}
  \totald{t} A_{e}(t) = - \frac{1}{2} (\gamma + \kappa) A_e(t)
    - i \sqrt{\frac{\kappa}{\tau}}.
\end{equation}
Solving this equation with $\gamma =0$, using the initial
condition $A_e(0)= 0$, will give the same result as in Eq.\
(\ref{E:NoSpont}).

Before considering the solution to this differential equation,
consider the slightly altered situation in which the field starts
in the vacuum state, and the atom in the excited state.  In this
case it is easy to see that the second term on the right side of
Eq.\ (\ref{E:SpontDiff}) will vanish. Then the equation becomes,
\begin{equation}
  \totald{t} A_e (t) = - \frac{1}{2} (\gamma + \kappa) A_e(t),
\end{equation}
with the initial condition $A_e(0) = 1$.  This must give the
standard exponential decay due to spontaneous emission in the
vacuum
\begin{equation}
  A_e(t) = e^{- \frac{1}{2} \Gamma t}.
\end{equation}
This allows us to equate $\Gamma = \gamma + \kappa$, where
$\kappa$ is again seen to be the spontaneous emission rate due to
the contribution of the modes in the paraxial beam.

\begin{figure}[t]
\scalebox{.45}{\includegraphics{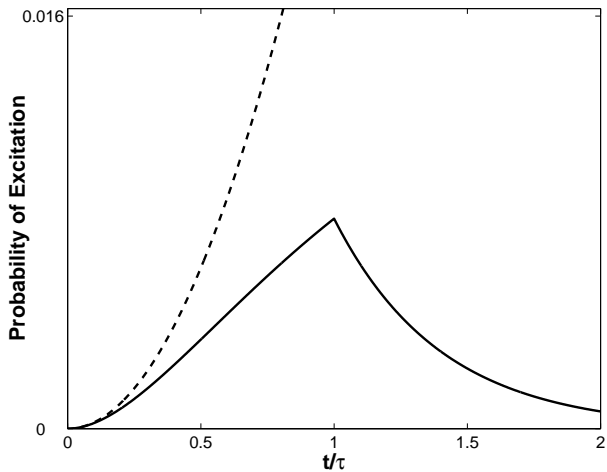}}
\caption{\label{F:Singlephoton}The probability for a single photon
in free space to excite a two-level atom (solid line) differs
significantly from the standard Rabi flopping solution, in which
an atom is coupled to a photon in a single-mode cavity ( dashed
line).  The parameters used here are $\Gamma \tau =2.5$, and
$\kappa/\Gamma = 1/50$ (see text).}
\end{figure}

The general solution to Eq.\ (\ref{E:SpontDiff}) is,
\begin{equation}\label{E:Excited}
  A_e (t) = - \frac{2i}{\Gamma} \sqrt{\frac{\kappa}{\tau}}
    \left( 1 - e^{-\Gamma t /2} \right) ,
\end{equation}
holding for times $0\leq t \leq \tau$.  The probability of the
atom being in the excited state is then,
\begin{equation}
  P_e(t) = 4 \frac{\kappa}{\Gamma^2 \tau}
    \left(1 - e^{-\Gamma t /2}\right)^2,
\end{equation}
during the same interval.  This is a monotonically increasing
function of $t$, and so achieves its maximum at the upper limit $t
= \tau$, after which the excitation probability can only
decay(Fig.\ \ref{F:Singlephoton}). Then the maximum probability
for any $\tau$ can be found by solving,
\begin{equation}
  \frac{d}{d\tau} P_e(\tau) = 0 \to \Gamma \tau \approx 2.5.
\end{equation}
At this point the probability is $P_e \approx .8 \kappa/\Gamma$,
which is necessarily less than 1 given that $\kappa/\Gamma <1$ as
previously discussed.  Thus, in the paraxial approxiamtion, no
single photon pulse can be constructed that excites an atom with
certainty.

This result is, of course, not surprising. Symmetry ensures that
the only single photon pulse capable of exciting an atom with unit
probability is the time reversal of a spontaneously emitted packet
\cite{Chan2002}.  Such a pulse represents the vector spherical
harmonic associated with the atomic radiation pattern (assuming
emission on a transition with well defined angular momentum), a
field not captured in the scalar paraxial approximation. Moreover,
even including the vector nature of the field beyond paraxial will
not be sufficient to yield high excitation probability.  The field
must be well ``mode matched'' to the the atom's radiation pattern
to give strong coupling between the atom and a single photon in
free space \cite{vanEnk2001b}.

\subsection{\label{s:Multiple}Interaction with a large $n$-Photon Fock Pulse}
In the last subsection we showed that in free space the coupling
of a single photon to an atom will not lead to coherent Rabi
oscillations. In contrast, for an $n$-photon Fock state with a
large $n$, we expect the system to be dominated by stimulated
emission.  A rigorous treatment of this probem using the Bethe-Ansatz was given by V.~I.~Rupasov and V.~I.~Yudson \cite{Rupasov1984}.  We show here how this phenomenon is recovered simply in the
present formalism.

In general, an arbitrary division may be envisioned in Hilbert
space that separates system from reservoir.  In Sec.\
\ref{S:Propagation} the system was chosen to be the paraxial field
modes plus the atom. Alternatively we may take the system to
consist of  the atom interacting with the single pulse mode
defined by creation operator $\hat a^\dag [f]$ in Eq.\
(\ref{E:SingleInit}), with all other modes making up the
environment.  Such a choice can always be made, and cannot change
the physics.  We may then choose to ignore all environmental modes
if $\Gamma \tau \ll 1$, with the caveat that any effects that occur in
the system must be have scales much larger than $\Gamma \tau$
in order to be considered valid.  

Consider a pulse of length $\tau$ such that
$\Gamma \tau \ll 1$.  Further, take the state of the system to
only have excitations in this pulse mode such that,
\begin{equation}
  \ket{\psi (0)}= \frac{1}{\sqrt{n}!} \left( \hat{a}^{\dag}[f] \right)^n \otimes \ket{0,g} =\ket{n,g}.
\end{equation}
The rest of the modes are in the vacuum state and are treated as
an environment. Ignoring terms of order $\Gamma \tau$ the system
evolves under the single mode Hamiltonian
\begin{equation}
  \hat{H} = \hbar g_{\text{eff}}\left( \hat{a}[f] \hat \sigma_{+} + \hat{a}^\dag [f] \hat\sigma_- \right),
\end{equation}
This is none other that the Jaynes-Cummings Hamiltonian restricted
to an initial manifold with $n$ excitations and having effective
coupling constant $\hbar g_{\text{eff}} =d \sqrt{2 \pi \hbar
\omega_0/ V_{\text{pulse}}}$ , where $V_{\text{pulse}}=Ac \tau$ is
the pulse volume. The system undergoes the familiar coherent Rabi
flopping within the two dimensional manifold, as in Eq.\
(\ref{E:Oscillation}),
\begin{equation}
  \ket{\psi (t)}=\cos(g_{\text{eff}}\sqrt{n} t)\,\ket{g}\ket{n}
    - i\sin(g_{\text{eff}} \sqrt{n} t)\,\ket{e}\ket{n-1}.
\end{equation}

This solution applies to the single photon case as well.   The
probability amplitude from Eq.\ (\ref{E:Excited}) limits to,
\begin{equation}
  A_e(t) \to -i g_{\text{eff}} t \left(1 + O(\Gamma t)\right),
\end{equation}
which is the correct limit of sinusoidal Rabi oscillation. However
during the pulse duration not even a single oscillation can occur since
$g_{\text{eff}} = \sqrt{\kappa / \tau} \ll  1/ \tau$ . Thus, for
true oscillations to occur, one must have
\begin{equation}
    g_{\text{eff}}  \sqrt{n} \gtrapprox \frac{1}{\tau}
    \quad \text{or} \quad n \Gamma \tau \gtrapprox \frac{A}{\sigma_{\text{eff}}} .
\end{equation}
The last inequality can be interpreted as saying that the mean
number of photons emitted via stimulated emission into the pulse must
dominate over spontaneous emission, even when the spontaneous
photons are paraxial.  When these conditions hold we may \textit{consistently} ignore all initially unoccupied modes, yet still recover dynamics in agreement with the usual Jaynes-Cummings Hamiltonian.  The multimode description of the field becomes superfluous.

Finally, what is the measurement strength associated with probing
an atom using a large photon number Fock state pulse?  Unlike the
coherent state case, the field does not factorize into
uncorrelated temporal slices.  In fact, when viewed in terms of
the coarse grained modes, the Fock pulse in highly
\textit{entangled}. Detection of photons at the leading edge of
the pulse will introduce new fluctuations in the trailing edge,
which has yet to interact with the atom.  This implies that the
usual Markov approximations do not hold.  The Fock pulse is most
naturally treated as part of the ``system'' rather than an
``environment'' which continuously carries information away from
the atom.

\section{\label{S:Faraday} QND Measurement via Faraday Polarization Spectroscopy}

The resonant interaction considered up to this point, though
fundamental in nature, has little practical application to the
problem of continuous measurement in free space since the
measurement strength is always bounded from above by the
spontaneous emission rate.  We thus turn our attention to an
off-resonant interaction.  In particular, we consider the problem
of measuring a spin component of an atom through the Faraday
effect wherein the linear polarization of a probe laser rotates by
an amount proportional to the magnetization of the sample.  This
interaction provides for a ``quantum nondemolition measurement''
(QND) of the atom, i.e.\ the probe measures an observable (here
the atomic spin along the laser propagation direction) without
perturbing that observable \cite{Kuzmich1998, Takahashi1999}. Such
an interaction has been applied to ensembles of atoms to produce
spin squeezed states \cite{Kuzmich2000} and to demonstrate
entanglement between two spatially separated ensembles
\cite{Julsgaard2001}.

We review here the basic physics associated with this measurement
scheme. The interaction energy is that of an electric-polarizable
particle in a vector field whose Hamiltonian may be written as,
\begin{equation}\label{E:FarHam}
  \hat{H}_\text{\text{int}} = -\hat{\vect{E}}^{(-) } \cdot \hat{\overset{\leftrightarrow}{\alpha}} \cdot \hat{\vect{E}}^{(+)},
\end{equation}
where   $\hat{\overset{\leftrightarrow}{\alpha}}$ is the atomic
polarizability tensor and $\vect{E}$  is the complex electric
field amplitude. Expressing this equation in terms of irreducible
tensor components, the interaction can be decomposed into an
effective scalar, vector, and symmetric rank-2 contribution.  We
consider here the case of alkali atoms probed on the so-called
$D2$ line $S_{1/2} \to P_{3/2}$, for which the ground state atomic
polarizability operator is \cite{Deutsch1998}
\begin{equation}
\overset{\leftrightarrow}{\alpha}=
    \alpha_{\text{lin}} \left(\hat{1}+\frac{1}{2}\left(\vect{e}_{+} \vect{e}_{+}^* - \vect{e}_{-} \vect{e}_{-}^* \right) \hat{\sigma}_z \right) .
\end{equation}
Here $\vect{e}_{\pm}$ are the right and left helicity vectors
relative to the quantization axis along the probe propagation
direction,  $\hat{\sigma}_z = \ket{\uparrow}\bra{\uparrow} -
\ket{\downarrow}\bra{\downarrow}$ is the Pauli spin operator for
the ground state electron, and $\alpha_{\text{lin}}$ is the atomic
polarizability for fields with linear polarization.   This
equation has a simple interpretation.  The first ``scalar term''
gives rise to an effect that depends solely on the field intensity
whereas the term proportional to  $\hat{\sigma}_z$ depends on the
field elipticity.  Clearly the irreducible  rank-2 component
vanishes here  \cite{Happer1967, Deutsch1998, BiggerF}.

Under the paraxial approximation we take the field to be
approximately a plane wave with two polarization vectors.  The
quantum field associated with the complex amplitude is
\begin{equation}
    \hat{\vect{E}}^{(+)} = \sqrt{\frac{2 \pi \hbar \omega}{V}} \left( \hat{a}_{-}  \vect{e}_{-} + \hat{a}_{+} \vect{e}_{+} \right) e^{ikz},
\end{equation}
where $V$ is the effective quantization volume for the propagating
mode.  Substituting this into Eq. (\ref{E:FarHam}), the quantum
Hamiltonian becomes,
\begin{equation}
\hat{H}_{\text{int}} = -\frac{2 \pi \alpha_{\text{lin}}}{V} \hbar
\omega \left[ \left( \hat{N}_{+}  + \hat{N}_{-} \right) +
\frac{1}{2} \left( \hat{N}_{+}  - \hat{N}_{-} \right)
\hat{\sigma}_z \right],
\end{equation}
where $\hat N_\pm$ is the number operator for photons in the
positive or negative helicity states.  The scalar term gives rise
to an overall phase shift (index of refraction) and thus can be
absorbed into the free field Hamiltonian.  The vector term gives
rise to the Faraday effect. Recognizing $ \hat{J}_z  = (
\hat{N}_{+}  - \hat{N}_{-} )/2 $ as the total field helicity, the
effective interaction Hamiltonian takes the QND form,
\begin{equation}
  \hat{H}_{\text{int}} = -\frac{2 \pi \alpha_{\text{lin}}}{V} \hbar \omega \hat{J}_z \hat{\sigma}_z .
\end{equation}
Under this Hamiltonian the photon spin becomes correlated with the
atom's magnetic moment and thus the laser polarization may act as
a meter for the atomic spin.  Since $\hat{\sigma}_z$ also commutes
with the system-meter Hamiltonian it is clearly a QND variable
\cite{Scully1997}.

Our Hamiltonian still contains the undefined quantization volume
$V$.  To rectify this we follow the formalism introduced in Sec.\
\ref{S:Propagation}.  Introducing propagating modes of duration
$\Delta t$ so that $V \rightarrow A c \Delta t$ the Hamiltonian
becomes,
\begin{equation}
  \hat H_{\text{int}} = - \sum_i 2\frac{\hbar \chi}{\Delta t} \Theta_i(t) \hat{J}_z \hat{\sigma}_z ,
\end{equation}
where $\chi = \pi \alpha_{\text{lin}} \omega / (c A) \approx
(\sigma_0 / A) [\Gamma / (-2 \Delta)]$.  Here $\Delta$ is the
(far)detuning from the atomic resonance with  linewidth $\Gamma$,
and  $\sigma_0$ is the on resonance absorption cross-section for
linear polarization.

With the Hamiltonian so defined, the evolution of the system may
be calculated.  The system shall consist of an atom interacting
with a laser beam that is linearly polarized in the $x$-direction.
The initial state of the laser is then,
\begin{equation}
  \ket{\Phi}_\text{probe} = \otimes_i\frac{1}{\sqrt{2}}
     \left(\ket{\alpha}_{i x} \ket{0}_{iy}\right),
\end{equation}
where $x$ and $y$ label the two orthogonal linear polarization
modes.  As in Sec.\ (\ref{s:Laser}) $|\alpha|^2 = P \Delta t /
\hbar \omega_0$, with $P$ the power in the beam. Given this
initial state for the field, we may define the marginal density
operator for the atom alone by tracing over the field at time $t$.
Since the Hamiltonian (\ref{E:FarHam}) only couples the atom to
the $k$'th field mode in time interval $(t_k ,t_k + \Delta t]$,
the reduced atomic state evolves during this interval as,
\begin{equation}\label{E:ReducedEvolution}
  \hat{\rho}(t_k + \Delta t) = \text{Tr}_{k}\left[
    \hat U_i \ket{\alpha}_{i x}\bra{\alpha}_{ix}
    \otimes \ket{0}_{i y}\bra{0}_{iy}\otimes \hat{\rho} (t_k) \hat U^\dag_i
  \right].
\end{equation}
In the interaction picture the unitary evolution becomes,
\begin{align}
  \hat{U}_i \ket{\alpha}_{i x} \ket{0}_{i y}
    &=e^{- i \chi \hat{\sigma}_z (\hat N_{+ i} - \hat N_{- i})} \ket{\alpha}_{i x} \ket{0}_{i y},\nonumber \\
    &= \ket{\alpha \cos (\chi \hat \sigma_z)}_{x i} \ket{-\alpha
      \sin(\chi \hat \sigma_z)}_{y i}.\label{E:Unitary}
\end{align}
This expression should be interpreted in the context of a matrix
element, given that atomic operators appear in the labels for the
field kets. Clearly, the field and atomic states will generally
become entangled by the interaction.

The map in Eq.\ (\ref{E:ReducedEvolution}) constitutes a
continuous measurement on the field. To see this explicitly, we
must expand in powers of $\Delta t$, and since $|\alpha|^2 \propto
\Delta t$, this is equivalent to an expansion of the state kets in
Eq.\ (\ref{E:Unitary}) in powers of $\alpha$. In the Fock (photon
number) basis we have,
\begin{multline}
  \hat{U}_i \ket{\alpha}_{i x} \ket{0}_{i y}\approx
  \left(1 - \frac{|\alpha|^2}{2}\right) \left[
    \ket{0}_{x i}\ket{0}_{y i}\right.\\
    \qquad+ \alpha \cos (\chi \hat \sigma_z) \ket{1}_{x i}\ket{0}_{y i} \\
    \left.- \alpha \sin(\chi \hat \sigma_z) \ket{0}_{x i} \ket{1}_{y i}
  \right]
\end{multline}
Substituting this back into Eq.\ (\ref{E:ReducedEvolution}) one
finds,
\begin{multline}
  \hat{\rho}(t_k + \Delta t) = \rho(t_k)
    + |\alpha|^2 \left[ -\rho(t_k)\right.\\
    + \sin(\chi \hat \sigma_z)\rho(t_k) \sin(\chi \hat\sigma_z)\\
    + \left.\cos(\chi \hat \sigma_z) \rho(t_k)
\cos(\chi \hat \sigma_z)\right].
\end{multline}
This form may be further simplified since $\chi \ll 1$ for a
single atom probed by a far off resonance laser.  Taking the
limit, $\Delta t \to 0$, and keeping terms to second order in
$\chi$ gives the master equation,
\begin{align}\label{E:FarMaster}
  \frac{d \hat{\rho}}{dt} &=
    \frac{P \chi^2}{\hbar \omega_0} \left[ \hat \sigma_z \hat{\rho} \hat \sigma_z
    - \frac{1}{2} \{\hat \sigma_z^2,\hat{\rho} \} \right] \nonumber\\
    &= -  \frac{\kappa}{2}  \left[  \hat{\sigma}_z ,\left[\hat{\sigma}_z, \hat{\rho}  \right] \right] ,
\end{align}
where the measurement strength can be easily identified from the
familiar Linblad form of the mater equation \cite{Gardiner2000}
as,
\begin{equation}
  \kappa = \frac{P \chi^2}{\hbar \omega_0},
\end{equation}
Note that the steady state solutions to Eq.\ (\ref{E:FarMaster}) are the 
Dicke states, so this technique provides a QND measurement of 
spin.  This expression was also derived by Thomsen and Wiseman in the
context of control of atom-laser coherence \cite{Thomsen2002}.

The physics of the continuous measurement under consideration here
differs substantially from that of the resonant Jaynes-Cummings
interaction studied in Sec.\ \ref{S:Interaction}. Here the
measurement strength depends upon the power in the beam, whereas
the previous result had no such dependence. For the resonant case,
$\kappa$ could be explained as arising from those spontaneous
photons emitted into paraxial modes.  In contrast, the dependence
of the QND measurement on the laser power indicates that the
measurement strength is due to the coherent redistribution of
photons between the polarization modes in a manner depending on
the atomic state.

\section{\label{S:Summary} Summary and Discussion}
Continuous quantum measurement, once studied only in gedanken
experiments, can now be realized in laboratory applications
\cite{Mabuchi1999, Reiner2001}. Such applications utilize the
ability to continuously gather information from a probe coupled to
a quantum system. The system state then evolves under a stochastic
master equation, characterized by the measurement strength.  In
this article we examined how one may derive the measurement
strength for paraxial laser probes in free space by considering
two examples.

We considered first the fundamental system of an atom coupled to a
laser beam. Classically the atom-laser interaction leads to Rabi
flopping which is often used to manipulate coherent superpositions
of atomic states.  In particular laser control pulses can be used
to implement quantum logic gates \cite{Wineland1998}. Quantum
mechanically continuous measurement of the atom by the quantum
laser pulse can lead to entanglement between the system and the
probe, inducing errors in logic gate operation. We found that this
is not a concern as the measurement strength is bounded from above
by the total spontaneous emission rate and can thus be neglected
for short interaction times. In fact, the measurement rate may be
attributed to spontaneous emission into the paraxial modes, which
will always be small compared to emission into $4\pi$ steradians
with the appropriate vector field. We also studied the distinction
between continuous observation by a coherent beam and by a pulse
with well defined photon number.  In the latter case correlations
between the quantum fluctuations at different points in the pulse
disallow a clean separation between system and probe.  Instead,
the entire atom-field system will coherently Rabi flop if the
number of photons is sufficiently large. For a single photon,
however, we showed that Rabi flopping cannot occur in free space.
Such a result is essential lest a distant detector be able to
instantly sense the presence of the atom. This highlights the need
to treat the propagation of quantum fluctuations in a traveling
wave geometry with great care to avoid contradictions with
causality.

Another mechanism allowing for continuous measurement of an atom
is the Faraday rotation of an off-resonant polarized beam, induced
by the atom's magnetic moment. This interaction produces a QND
measurement of the atomic spin. The measurement is due to
stimulated redistribution of photons between pairs of occupied
modes, in contrast to the resonant Rabi interaction where
measurement results from spontaneous emission. Since the
measurement strength is proportional to the laser intensity it can
be made much larger than the spontaneous emission rate. Of course
spontaneous emission is not the only source of noise.  In order to
see the effects of quantum back-action the quantum ``projection"
noise \cite{Itano1993} must be large compared all other noise
sources, such as those due to photo-detection. This implies that
an experiment in which the back-action is important must enhance
the coupling constant without increasing the shot-noise, possibly
through the use of an optical cavity or by maintaining a large
ensemble of atoms collectively coupled to the probe. The latter
approach has led to the observation of spin squeezing
\cite{Kuzmich2000} and ensemble entanglement \cite{Julsgaard2001}.
In such a situation it would be interesting to understand the
effective measurement strength by the probe on any one member of
the ensemble. Exploring this question will require us to consider
how quantum correlations are shared within a multipartite system
interacting with a probe.  We plan to address this question in
future work.

\begin{acknowledgments}
IHD acknowledges extremely useful discussions about the problem of
single photon Rabi flopping with Ken Brown, Bill Wootters, and
David DiVincenzo while in residence at the IPT, Fall 2001, with
support from the ITP's NSF Grant No.~PHY99-07949.  We thank also
Poul Jessen for discussions relating to quantum projection noise
and measurement back-action.  IHD and AS were supported under the
auspices of NSF Grant No.~PHY-009569 and the ONR Grant
No.~N00014-00-1-057.
\end{acknowledgments}


\begin{thebibliography}{99}
\bibitem{Coherence}
In general, the atom volume of interest is its ``coherence
volume''. For simplicty, we treat here the case of a pure atomic
wave packet, with no loss of generality.
\bibitem{BiggerF}
This result generalizes under the assumption that the probe
detuning is large compared to the hyperfine splitting in the
$P_{3/2}$ manifold. We can use the Lande projection theorem to
make the substitution $\hat{\sigma}_z \Rightarrow \hat{\Sigma}_z
\equiv \hat{F}_z/F$ where $F$ is the total ground-state angular
momentum including nuclear spin.
\end{thebibliography}
\end{document}